# Direct Evidence of Tungsten Clustering in $W_{0.02}V_{0.98}O_2$ Thin Films and its Effect on the Metal-to-Insulator Transition

Xiaoyan Li,[†,‡,¶] Alexandre Gloter,[‡,*] Alberto Zobelli,[‡] Hui Gu,[†] Xun Cao,[†] Ping Jin,[†] Christian Colliex[‡]

[†] State Key Laboratory of High Performance Ceramics and Superfine Microstructures, Shanghai Institute of Ceramics, Chinese Academy of Sciences, Shanghai 200050, China

[‡] Laboratoire de Physique des Solides, CNRS UMR 8502, Université Paris Sud 11, Orsay 91405, France

[¶] University of Chinese Academy of Sciences, Beijing 100049, China

KEYWORDS: vanadium dioxide thin film, tungsten dopant, cluster, scanning transmission electron microscopy, metal-to-insulator transition, ab-initio

**ABSTRACT:** Substitutional tungsten doping of $VO_2$ thin films and its effect on the metal-to-insulator transition are investigated by means of X-ray diffraction, Cs-corrected scanning transmission electron microscope (STEM) and *ab-initio* simulations. The $W_{0.02}V_{0.98}O_2$ thin films deposited on (001) sapphire are studied in both planar and transverse geometries. The tungsten atoms are distinguishable from the V atoms in the Z-sensitive high angle annular dark field (HAADF) STEM image and their nature is further confirmed by electron energy loss spectroscopy (EELS). The W dopants are found to substitute in the V sites and form local clusters at first neighbor, preferentially along the $<010>_R$ directions. *Ab-initio* modeling for this 2 at.% W doped $VO_2$ confirm the experimentally found W clustering mechanism to be the most stable substitutional configuration and they demonstrate that the binding energy of such cluster is 0.18 eV. Driving forces for short range ordering are also obtained along the $<011>_R$ and $<110>_R$ directions. Strong energetic penalty is found for the $<001>_R$ direction. Simulations indicate that the clustering helps in stabilizing the tetragonal structure, while a diluted W dopant induces more structural distortion and V-V pairing. This suggests that the clustering mechanism plays a critical role in the transition temperature evolution with the W dopants.

## INTRODUCTION

Vanadium dioxide ($VO_2$) undergoes a first-order metal-to-insulator transition (MIT) at ca. 340 K from a high-temperature rutile tetragonal metallic phase ($P4_2/mnm$, denoted as the R phase) to a low-temperature monoclinic insulating phase ($P2_1/c$, denoted as the M1 phase), characterized by a pairing and tilting of the V-V chains along the *c* axis of the R phase (see Figure 1).[1] In its lower symmetry M1 phase, the formation of the V-V dimers and the doubling of the unit cell are coupled with a splitting of the $a_{1g}$ ($d_∥$) electron band and an opening of a small band gap.[2,3] Associated with this transition, a dramatic change in both electron resistivity and near-infrared transparency are observed which may offer various potential applications such as thermochromic 'smart window' coatings,[4-6] multistate memory devices,[7,8] and ultrafast switches[9] etc. Thus, many studies have attempted to tune the MIT properties,[10,11] among which tungsten doping has received particular attention when compared to other dopants such as Ru, Nb and Mo. Indeed, W dopants efficiently lower the transition temperature to room temperature by a rate of ca. 20-28 K/at.% for the bulk and film [2,12,13] and of ca. 50 K/at.% for the nano structures.[14,15]

The mechanism behind this MIT transition has been debated for several decades and it has attracted great attention from various research areas. Two different points of views, as being more a Mott-type transition characterized by strong electron-electron correlations[16,17], or more a Peierls-type transition triggered by the structural V dimerization, have been supported experimentally and theoretically.[2,18,19] More recently, some studies taking into account both the structural pairing and the strong coulomb interaction proposed a correlation-assisted Peierls model.[3,20-22] In the case of the W-doped $VO_2$ system, disputes extend to the respective role of the W dopants as introducing extra electrons or giving rise to internal structural effects. On one hand, the dopant W ion has been demonstrated to have a 6+ valence which, according to the charge compensation, results in the formation of two $V^{3+}$ per dopant.[23,24] Furthermore, photoemission measurements also revealed that the spectral weight at the Fermi level is drastically changed with this electron doping, suggesting a reduction in the effective Coulomb repulsion. This strongly disturbs the insulating half-filled $V^{4+}$-$V^{4+}$ ($d^1$-$d^1$) electron band configuration and favors the existence of the metallic phase at a lower temperature.[23,25,26] On the other hand, since an external stress can tune the MIT transition temperature efficiently by ca. 10 K/GPa,[27,28] the internal stress due to the W dopant is also expected to act similarly.[29-34] The analysis of the local structures around both V and W dopant has been performed by several groups,[29-33] mostly by synchrotron Extended X-Ray Absorption Fine Structure (EXAFS) method. These works report that the $VO_2$ local structure is strongly altered by the W dopants: in the M1 phase, W dopants drive the de-twisting and decoupling of their nearby V-V pairs, increase the symmetry toward the R structure and act as nucleation centers to facilitate the structure transition;

otherwise in the R phase, the W dopants induce the dimerization of the V-V chains. Corresponding first principle calculations also suggest that these structural alterations can reduce the energy difference between the two phases and therefore decrease the transition temperature.[33, 34]

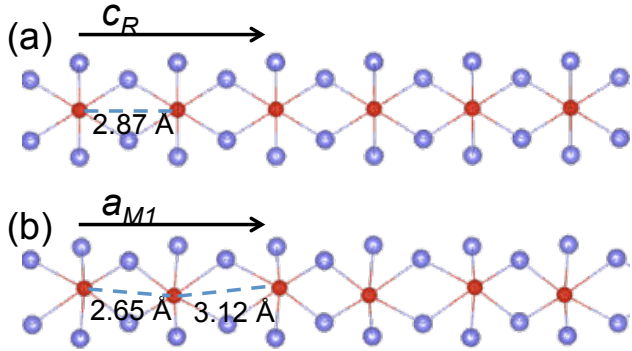

Figure 1. Layout of the V-V chains along (a) the $c$ axis of the R phase and (b) the corresponding $a$ axis of the M1 phase. The two phases are related to each other by $a_{M1} = 2c_R$, $b_{M1} = a_R$, $c_{M1} = a_R - c_R$. Red and blue balls denote vanadium and oxygen respectively.

In this article, we report the W doping mechanism in $VO_2$ thin films as observed mainly by the atomic-resolution high angle annular dark field (HAADF) STEM imaging where W atoms exhibit a higher intensity compared to the V atoms. The electron energy loss spectrum (EELS) analysis further indicates that the W atoms. By observing the samples from several directions (both planar and cross-section view), we have systematically found alternated W-rich and W-poor planes on the $\{020\}_R$ plane family. Such an ordered structure indicates W clustering, and in particular along $<010>_R$ orientations. These clustering trends are further confirmed by *ab-initio* simulations of the 2 at.% W doped $VO_2$, which find the most energetically favored W-W distances to be 4.23 Å along $<010>_R$. Other preferences for clustering are found as well for $<011>_R$ and $<110>_R$ orientations. The influence of such W clustering in the $VO_2$ tetragonal structure and especially in the V-V chains nearby the dopants are also studied based on the calculated relaxed supercell structures. A diluted W dopant is found to disturb strongly the $VO_2$ tetragonal feature, and form more V-V pairs. However, the clustering of W dopants is shown to contribute in reducing and localizing the structural distortion, which may help in stabilizing the metallic phase and thus play a crucial role in reducing the transition temperature.

## EXPERIMENTAL SECTION

Epitaxial $VO_2$ films with 2 at.% W dopant were deposited on (001) sapphire by reactive magnetron sputtering. A substrate temperature up to 450 °C with a deposition time of 200 min was applied in the epitaxial growth (see supporting information for more film deposition details and Figure S1).[12, 35] Planar view TEM samples were prepared by normal tripod polishing and followed by ion milling process only from the substrate side, in such a way that only the $VO_2$ thin film was obtained in the electron transparent thin area. Cross-section TEM samples were prepared by gluing the sample face to face by G2 (GATAN Co.), and then following the conventional specimen preparation techniques, including mechanical cutting, polishing and ion milling. The epitaxial structure of the deposited films was characterized by XRD with $\theta$-$2\theta$ scan (Ultima IV. Rigaku Co., Japan) and the MIT characteristic was investigated by measuring the electrical resistivity as a function of temperature. The detailed atomic structures of the W-doped $VO_2$ thin film were investigated by means of electron energy loss spectroscopy (EELS), atomic resolved bright field (BF) and high angle annual dark field (HAADF) imaging in a Cs-corrected Scanning Transmission Electron Microscope (STEM) (Nion Co., USA).

## RESULTS AND DISCUSSION

**Microstructures and MIT Properties of the Thin Films.** The main studied $VO_2$ thin films are about 70 nm thick grown on (001) sapphire, and they are composed of parallel column-like grains which show in-plane elongated shapes with nano scale dimensions (length 7-20 nm, width 2-10 nm, aspect ratio 1-5, approximately). In the XRD spectrum (see Figure S2 in the supporting information), the doped $VO_2$ thin film is observed to be R phase at the room temperature with the epitaxial relationship $(200)_R//(006)_S$ between the $VO_2$ and sapphire. Moreover with the indexation of the selected area electron diffraction (SAED) pattern of its planar TEM sample (inset of Figure S2 in the supporting information), an in-plane three-fold twinning structure is found and can be written as $(020)_R//\{110\}_S$. The microstructure of this W-doped polycrystalline $VO_2$ thin film is analogous to its undoped one which has been reported previously,[36] only with the grains slightly smaller in average size.

Resistivity curves as a function of the temperature for this 2 at.% W doped $VO_2$ ($W_{0.02}V_{0.98}O_2$) thin film in comparison with an undoped one are shown in the supporting information Figure S3. The phase transition temperature ($T_t$) is found to be ca. 347 K for the undoped sample which is comparable with that of a $VO_2$ single crystal, and 240 K for the doped sample suggesting a transition temperature reducing efficiency to be ca. 50 K/at.%. Moreover for the doped sample, the width of the hysteresis ($\Delta H$) is found to be ca. 20 K and the sharpness ($\Delta T$) of the transition process to be ca. 30 K which are wider compared to the undoped thin film ($\Delta H \sim 8$ K and $\Delta T \sim 20$ K) and the $VO_2$ single crystal ($\Delta H \sim 2$ K and $\Delta T \sim 0.1$ K). These broadening of the transition properties are mainly due to the presence of a higher density of grain boundaries within the doped film which contains averagely smaller grains.

**EELS Chemical Mapping for W Dopants.** STEM-HAADF images of the W-doped $VO_2$ thin films are shown in Figure 2a and with larger field of view in Figure 3a, c, Figure 4 and Figure 5 revealing clearly a modulation of the brightness of the atomic columns. Such brighter columns are considered as the ones containing more W dopants (tungsten Z=74, vanadium Z=23). For the sake of comparison, the STEM-HAADF image of the undoped $VO_2$ thin film is also provided in the supporting information Figure S4b and it presents a more homogeneous contrast of the pure V columns. Furthermore, the presence of the W atoms is confirmed by the EELS chemical mapping. The W $O_{2,3}$ edges is located at ca. 36 eV and cannot be used for atomic mapping due to the delocalization of the spectroscopic signal for such low energies. In addition, the EELS signal for the W impurity is hardly extractable in this energy range where the plasmonic loss prevails. Consequently, Figure 2 shows the results of an EELS spectral acquisition in the energy range from 1000 to 2000 eV where the W $M_{4,5}$ edges are present (onset at 1810 eV). Due to the sensitivity of the $VO_2$ thin film to the electron beam, the EELS spectra has been acquired within a typical time of tens of ms per pixel resulting in a weak W signature (electron doses below $10^7$ e$^-$·Å$^{-2}$ are necessary to avoid damage in these samples). Figure 2a shows the corresponding HAADF image of the selected area, while Figure 2b shows the HAADF image obtained synchronously during the EELS acquisition with a probe step of ca. 0.07 nm. In both images, 2 atomic columns show a much higher intensity. The corresponding EELS chemical mappings, in the range of the W $M_{4,5}$ edge as well as the pre-edge 'background' which mostly contains a thickness information, are constructed and shown respectively in Figure 2c and d. Although the signal-to-noise ratio for W edges is low, the brighter atomic columns are shown to contain increased spectral intensity, compared to the pre-edge 'background' mapping which shows almost no contrast. The intensity profiles across the four atomic columns in both the W and the pre-edge 'background' mapping are also shown respectively in Figure 2e and f, and they further confirm the existence of the W dopants in such brighter columns. It was not possible to quantify the W/V ratio with this EELS experiment because of the large energy difference between the V $L_{2,3}$ and the W $M_{4,5}$ edges. The positions of these W contained columns coincide with the brighter ones in HAADF images, which confirms that the HAADF technique is an efficient method to distinguish the W dopants and to investigate the atomic structure of these W doped $VO_2$ thin films. In addition, EELS measurements in the low energy range indicate that the HAADF images presented in this work correspond typically to thicknesses of 10 to 20 nm, which are roughly 25-50 vanadium atoms per column depending on the orientations (see next section). A random distribution of 2 at.% W dopants would then result in a typical average of ca. 1 W atom per column in the absence of any special ordered pattern.

EELS mapping for the pre-edge 'background'. (e and f) The intensity profiles cross the boxes in the EELS maps in (c and d) respectively. The arrow indicates the position of the atomic column with W dopant(s).

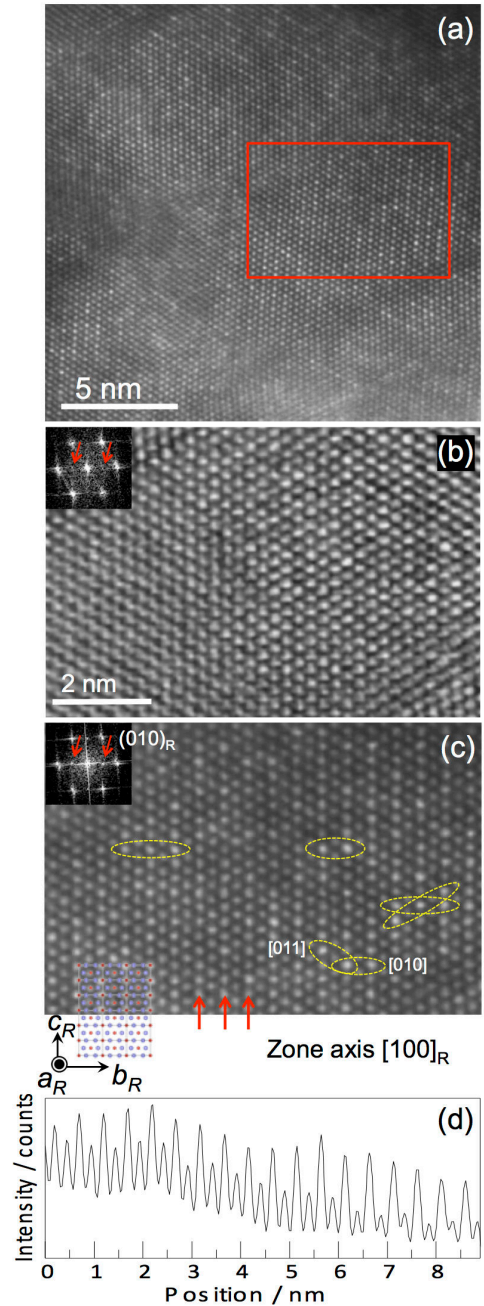

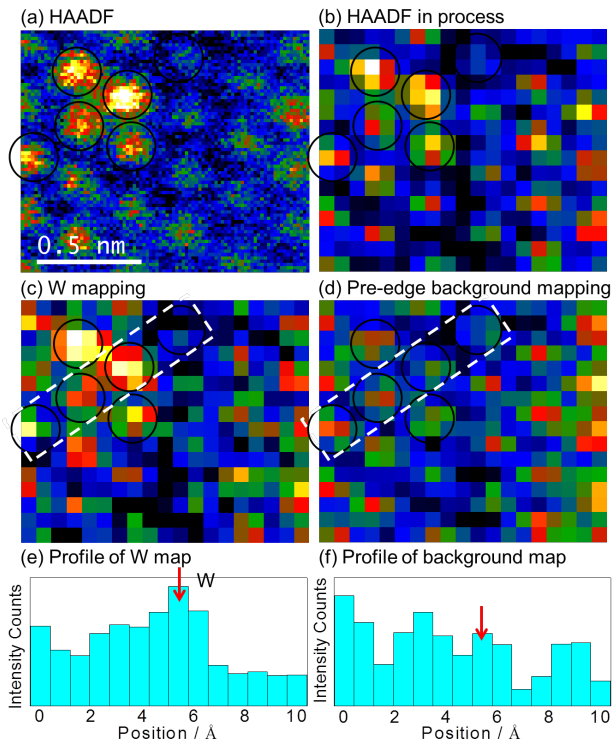

Figure 2. (a) HAADF image of the planar $W_{0.02}V_{0.98}O_2$ thin film TEM sample. (b) HAADF image taken synchronously with the recording of the EELS spectrum image. (c) EELS mapping for W $M_{4,5}$ edges. (d)

Figure 3. (a) HAADF image of the planar $W_{0.02}V_{0.98}O_2$ thin film TEM sample at zone axis $[100]_R$, containing several different grains. Enlarged (b) STEM-BF and (c) HAADF images of the boxed grain in (a): the insets are the FFTs of the images, and the corresponding crystal structure of the R phase is also fitted at the lower side of (c); some of the W rich planes are points out by arrows, and some examples of the W clusters are indicated by circles. (d) The intensity profile of (c) along the $b_R$ axis.

**STEM-HAADF Observation of W Clusters.** Figure 3a shows the STEM-HAADF image of the planar TEM sample of the $W_{0.02}V_{0.98}O_2$ thin film (in the R phase). Several grains with different in-plane orientations but sharing the same zone axis $[100]_R$ (the growth direction) can be seen in the figure. All grains show the same periodic ordered patterns: the darker and brighter planes are arranged alternately on the $\{020\}_R$ planes which correspond respectively to the W-poor and W-rich planes. The STEM-BF and HAADF images for one of these grains boxed in Figure 3a are enlarged in Figure 3b and c. The ordering of the W substitutions across the $\{010\}_R$ planes gives rise to the corresponding diffraction spots in the FFT of the HAADF image which are forbidden in the R phase, while the same spots are hardly visible in the FFT of the non-Z-sensitive BF images (see insets in Figure 3b, c). The intensity profile of this HAADF image along the $[010]_R$ direction is also given in Figure 3d, and it confirms the alternately W-rich planes in the HAADF images. Despite these overall ordered features, local inhomogeneities can also be found in the images, where several brighter columns stay adjacently with each other forming areas of few nm with higher density of interacting W dopants. Indeed, the relatively small depth of field occurring in a Cs-corrected STEM-HAADF is an advantage to visualize dopants distribution, and notably its clustering behavior,[37,38] since the contrast of dopants are enhanced significantly within a narrow depth around the focal plane. Consequently, the as-observed adjacent bright columns with similar intensities are indicative of W atoms located at similar depths (see Figure S5 for HAADF contrast analysis). Thus, such ordered structure demonstrates that the W dopants are not isolated and randomly distributed but correlated with each other in a specific manner suggesting a clustering. This clustering behavior can also be seen at a lower doping level such as 1 at.% W doped $VO_2$ thin films. In Figure 4a, a grain from such a thin film with the same crystallographic orientation as the one in Figure 3c is displayed, where some areas of several nm² exhibit very flat low contrast corresponding to regions with few W atoms while other areas show brighter W chains mainly with $[010]_R$ ordering revealing the formation of small local clusters of W atoms (see Figure 4b and c respectively).

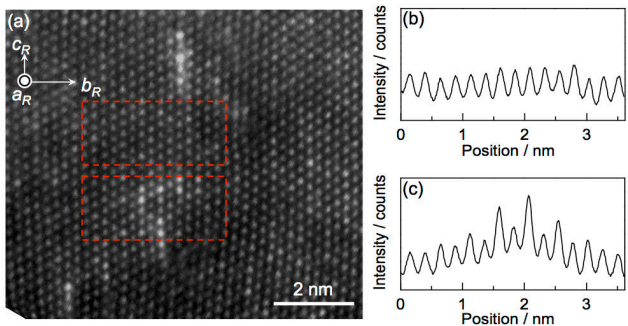

Figure 4. (a) HAADF image of the planar $W_{0.01}V_{0.99}O_2$ thin film TEM sample at zone axis $[100]_R$. (b and c) The intensity profiles respectively for the top and bottom boxed area in (a) along the $b_R$ axis.

In order to further figure out the favored clustering orientations for the W dopants, the cross section TEM samples were also investigated. Due to its polycrystalline in-plane three-fold twinning structure, four different typical zone axis can be reported: $[010]_R$, $[031]_R$, $[001]_R$, and $[011]_R$ (see Figure 5). They all share the same growth direction $[100]_R$ as indicated in the figure. Through all these different cross-section views, the HAADF images always show similar periodic ordered patterns on $\{200\}_R$ planes, indicating again the alternately ordered W-rich and W-poor planes (some of the W-rich planes are pointed out in the figures). According to the symmetry of the R phase, $[010]_R$ is equivalent to $[100]_R$ and the W ordering found in both the planar and cross-section views are coincident. The observation of alternating W-rich and W-poor planes, notably with spatial heterogeneities, along 5 directions (notably including $[100]_R$, $[010]_R$ and $[001]_R$ zone axis) also reinforce the occurrence of W clustering. Indeed, the formation of such long range ordering by distant W atoms without local clustering can be excluded, since it might give rise to modulation when projected along one zone axis but hardly along the 3 different orientations. In addition such distant W atoms configuration is not energetically favored as indicated by the *ab-initio* modeling in next section.

Several clustering trends can be obtained by observing these STEM-HAADF images along these different zone axes. First of all, since the W atoms prefer the second neighboring plane, *id est*, with a $\{020\}_R$ family modulation, it can be deduce that a clustering involving the W atoms distributed in a distance along $[½½½]_R$ orientation is probably unfavorable since it would result in an enhanced contrast in the first neighboring plane (See the $VO_2$ unit cell model in Figure 6). In a similar approach, the systematic observation of the intensity modulation across the $\{200\}_R = \{020\}_R$ plane family might indicates that W-W atomic spacing along $[100]_R$ and $[010]_R$ directions are favored. Such geometries are easily observed and have been indicated in several positions in Figure 3 and 5. Other possible atomic clustering modes such as $<101>_R$ and $<110>_R$ can also be frequently found in all the figures and have also been encircled for indicative purposes. In addition, the formation of bright column chains can also be observed along the $c$ axis such as in Figure 4, which could be interpreted either by W atoms preferential clustering along the $[001]_R$ within the first neighboring distance, or along the $<011>_R$ orientation. Nevertheless the $[001]_R$ clustering would result in randomly distributed very bright single columns when observed along the $[001]_R$ zone axis that are not observed in STEM-HAADF images (e.g., Figure 5c). In fact, further considering the intensity modulation of the bright atom columns along the $c$ axis which implying a minor difference of the W lying depths, $<011>_R$ cluster is in better agreement with the observations in all the figures. Moreover, the atomic spacing along $[001]_R$ in the edge-sharing $(V,W)O_6$ octahedral (ca. 2.8 Å) is much smaller than that of W atoms in the corner-sharing $WO_6$ octahedral in $WO_3$ structure (ca. 3.8 Å), which make the W $[001]_R$ clustering configuration structurally unfavored. Examples of such geometry and detailed explanation can also be seen in the Figure S6b and c.

According to these STEM-HAADF observations, the W dopants in the $VO_2$ are found to preferentially form an ordered substitutional structure by clustering along $<010>_R$ directions in association with $<011>_R$ and $<110>_R$. We have also checked that the similar W ordered cluster patterns can be observed as well in 2 at.% W doped $VO_2$ polycrystalline thin films deposited on a silicon substrate which are of more interest for electronic devices. In contrast to thin films grown on sapphire, these films do not exhibit epitaxial growth orientation or strain induced by the substrate (see Figure S7 in Supporting Information). Thus this clustering mechanism is concluded to be an intrinsic substitutional property for W dopants into $VO_2$ thin films.

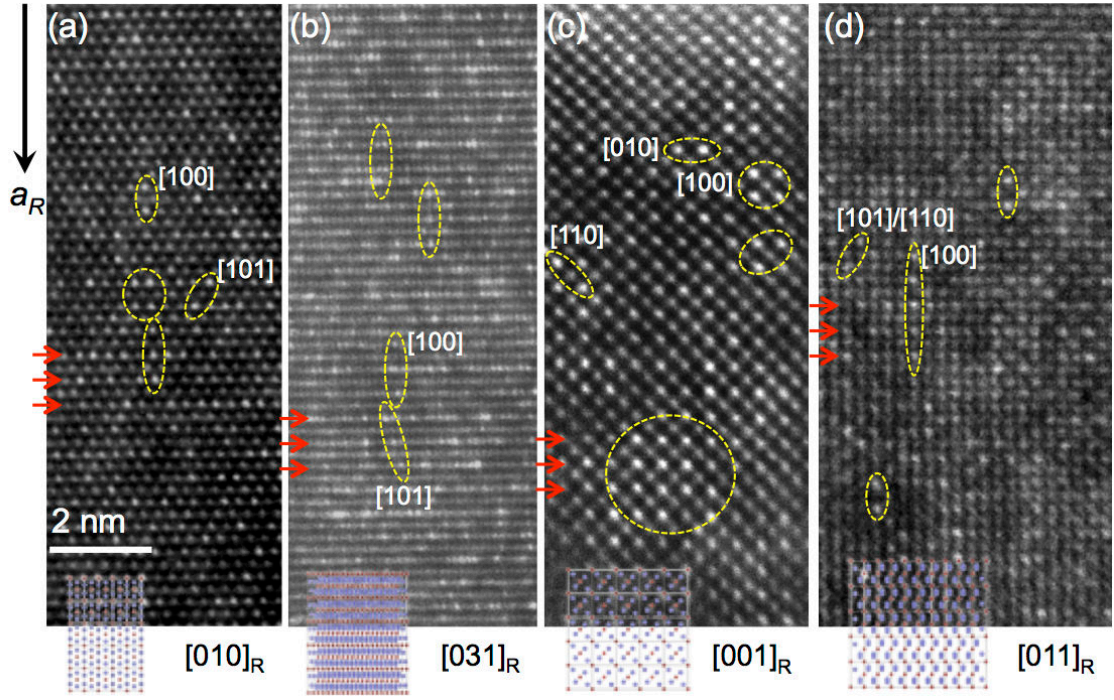

Figure 5. HAADF images of the cross-section $W_{0.02}V_{0.98}O_2$ thin film TEM samples at different zone axis: (a) $[010]_R$ zone axis, (b) $[031]_R$ zone axis, (c) $[001]_R$ zone axis, (d) $[011]_R$ zone axis. Their corresponding crystal structures are fitted in the lower part of the images; some of the W rich planes are pointed out by arrows, and some examples of the W dopant clusters are indicated by circles.

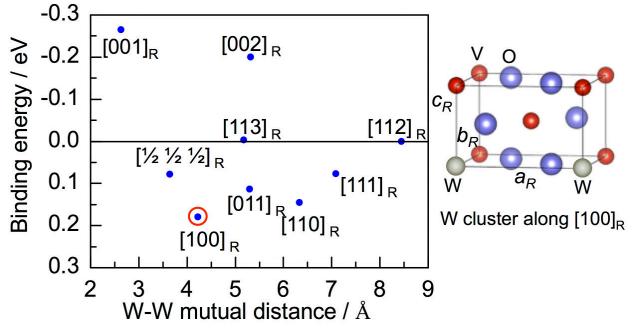

Figure 6. Binding energies as a function of the mutual distances between the two W dopants: the locating orientations of these two W atoms are also indicated next to their data (the two W atoms are first neighbor to each other, except in the $[002]_R$ and $[111]_R$ cases where they are second neighbor). The two W located along $[112]_R$ are the furthest W-W configuration we used, which is comparable to diluted limit and set as zero in the diagram. The corresponding VO2 structures showing the most stable W-W configuration along $[100]_R$ are also presented next to the diagram. The red, blue, and grey balls denote vanadium, oxygen, and tungsten respectively.

*Ab-initio* Modeling. *Ab-initio* modeling, in the framework of the density functional theory (DFT) as implemented in the AIMPRO code, has been performed to understand the W clustering mechanism in the VO₂ R phase and its effect on the MIT transition. Full structural relaxations have been obtained using the local density approximation (LDA) with an extended Gaussian basis set (40 Gaussians for W and V atom, 24 Gaussians for O atom) and a 3x3x5 supercell ($V_{90}O_{180}$, totally 270 atoms). The high temperature metallic R structure has been considered since its metallic character can be properly described within the LDA approximation. Furthermore, the thin film is deposited at a temperature where the R phase is the stable phase. Two V atoms in the supercell are replaced by W dopants (supercell $W_2V_{88}O_{180}$, 2.2 at.% W) to model the 2 at.% W doped VO₂. Nine possible W-W configurations are considered where the two W atoms are located along different directions with different mutual distances. The relative binding energies between the two W atoms as a function of their mutual distances are shown in Figure 6 where the locating orientation of the two W atoms is indicated next to the data point. The longest mutual distance between two W dopants is ca. 8.45 Å along $[112]_R$, which is comparable with the dilution limit. In the diagram, the structure corresponding to two W dopants located along $[100]_R$ is the most stable configuration with a 0.18 eV binding energy, and the VO₂ structure with such W-W configuration is shown aside. Moreover, for this W arrangement, the corresponding W-W distance (along $a_R$) decreases from 4.55 Å for the unrelaxed structure to about 4.23 Å after relaxation. Such W-W configuration is in well accordance with our experimentally found W favored clustering mode. Two other W-W configurations along $[110]_R$ and $[011]_R$, showing a slightly lower binding energy, are also frequently found in the HAADF images. Otherwise, when two W dopants lie in the V-V chains along $[001]_R$, either by first or second neighbors, the system is much unstable than the diluted one (negative binding energy < -0.2 eV). The binding energies of these different W-W clustering configurations are well consistent with the experimentally found W clustering trends.

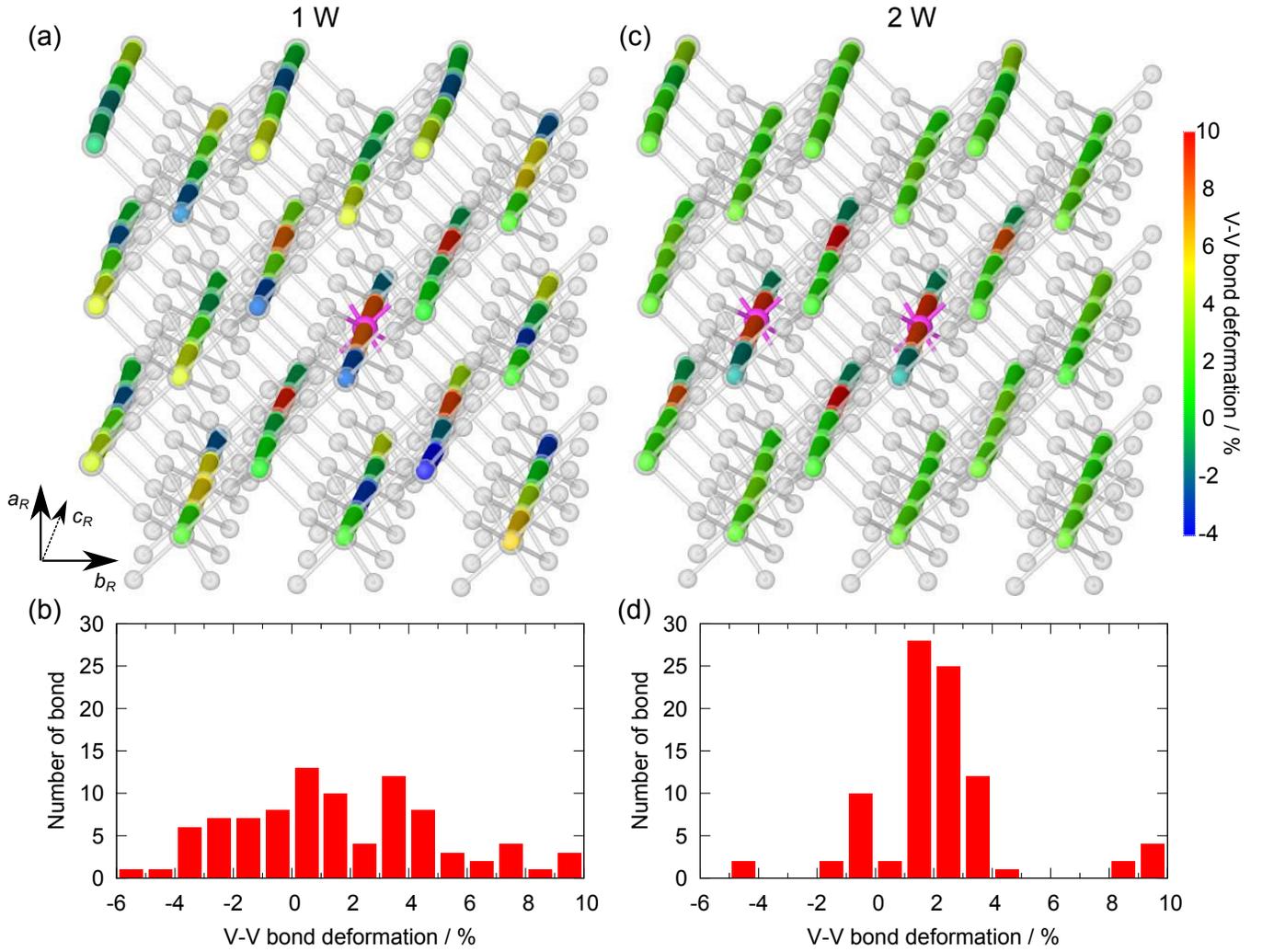

Figure 7. The relaxed VO$_2$ supercell for (a) one diluted W dopant and (c) two clustered W dopants. Solely, the V-V bonds in the V-V chains along the $c_R$ are colored according to their bond deformation whereas others bonds are transparent. The W atoms are set as pink. The V-V pairing can be seen where a long V-V bond (more red) and a short V-V bond (bluer) are located adjacently. The number of bonds as a function of the V-V (W-V) bond deformations is also shown below their corresponding structures in (b) and (d). The 0% bond deformation corresponds to 2.69 Å, the uniform V-V distance in a pure VO$_2$ after the structure relaxation, slightly smaller than the experimental value as expected for LDA based calculations.

The influence of this clustering mechanism on the MIT transition is also studied by comparing V-V distances along $c_R$ for the relaxed structure containing two W clustering along $<010>_R$, with those where a single diluted W atom is present. The supercells with color-mapped V-V distances along the V-V chains are shown in Figure 7a and c respectively for one diluted W doped and two-clustered W doped VO$_2$. The colors of the sticks correspond to the deformations of the V-V bond, as compared to the ones in a relaxed bulk VO$_2$. In this diagram, one can see that the uniform V-V chains of the R phase are distorted by substituting W atoms, some V-V bonds are elongated while others are shorten. In the diluted case, the W atom introduces a strongly structural perturbation through all over the supercell (even 1.2 nm away), and causes numerous V-V pairing (Figure 7a). However, when two W atoms are clustered, the structural distortion is more localized around the dopants, and the pairing of the V atoms is more regular and symmetric (Figure 7c). The V-V chains located at a distance further than ca. 0.78 nm away from the W-W cluster show almost no distortion, and the structure remains in a more uniform tetragonal feature. The distributions of the extremely elongated V-V bonds (denoted as red stick), which may play a critical role in inducing the V-V paring and thus opening the band gap, are also very different for both cases. The diluted W dopant introduces 6 distant V-V bonds, which are found with length above 2.87 Å (deformation > 7%) while there are only 4 of such V-V bonds per dopant for the clustered case (the Figure 7c shows totally 8 distant V-V bonds for the two clustered W case, which gives 4 V-V bonds/dopant). The distributions of the V-V bond deformations are summarized respectively in Figure 7b and d below their corresponding structures. The diluted W dopant makes the distribution of the V-V bond deformation to be strongly scattered all over its range (average at ca. 1.5% (2.73 Å) and the standard deviation is 3.4% (0.09 Å)). While in the clustered case, the distribution are more concentrate and homogenously distributed between 1% and 4% (average at 2.2% (2.75 Å) and the standard deviation is 2.8% (0.07 Å)) which are closer to an uniform tetragonal feature. In both cases, the average bond lengths are higher than the ones in a relaxed bulk VO$_2$, 2.69 Å, moreover, the volume increases for the one W and the two W contained supercell are of ca.

0.07% and 0.5% respectively which is in agreement with the previously XRD report.[39]

It has already been reported that the introduction of the W dopants into the R phase of $VO_2$ can induce a structural distortion and notably a V-V pairing in the chain where the W is located, but mostly the W dopants were considered in a diluted configuration.[29, 30, 34] Our HAADF observations and *ab-initio* simulations demonstrate the existing of the W clustering mechanism, which reduces the structural distortion effect, making it more localized around the dopant, while maintaining the further area in a better tetragonal feature. This structural effect might have implication on the MIT transition. Indeed the W dopant is the most efficient one to reduce the transition temperature (up to 30 K/at.%), compared to various other dopants like $Ru^{4+}$ (10 K/at.%), $Nb^{5+}$ (7.8 K/at.%), or even other hexavalent ions such as $Mo^{6+}$ (8 K/at.%)[2]. The formation of the $W^{6+}$ clusters may be an efficient balance between the structural distortion and electron doping effects in a correlation-assisted Peierls transition,[3, 22] where it introduces a large amount of electrons while maintaining a rather tetragonal feature. In particular, it is noteworthy that the W and Mo dopants introduce the same numbers of electrons while the latter is much less efficient in lowering the transition temperature (to our knowledge, no Mo clustering geometry has yet been reported in any $VO_2$ structures). Furthermore, these findings may also shed light in a new strategy for optimizing the MIT transition temperature: heterogeneous doping of thin films can now be aimed in a controlled manner to maximize the doping effect.

## CONCLUSION

The substitutional mechanism of W dopants into $VO_2$ thin films has been investigated mainly by atomic resolution HAADF imaging in a Cs-corrected STEM. The W doping distributions are observed for both planar and transverse geometries. It has been found that they preferentially form clusters, notably along the $<010>_R$ directions. *Ab-initio* modeling for such 2 at.% W doped $VO_2$ confirms such experimentally reported W clustering orientation to be the most energetically favored. Driving forces for short range ordering are also obtained along the $<011>_R$ and $<110>_R$ directions. The simulations further suggest that the clustering may be an effective mechanism to localize and reduce the W-induced structural distortions in the V-V chains. Therefore, beside the extra electrons introduced by the W ions, this clustering mechanism contributes to maintain the tetragonal structure, which may produce an efficient balance between a large electron doping and a weak structural distortion effects. Consequently, it could explain why W dopant is more efficient than other elements to reduce the MIT temperature.

## ASSOCIATED CONTENT

### Supporting Information.
Thin film deposited method, XRD spectra, Resistivity measurements of the $W_{0.02}V_{0.98}O_2$ thin films, BF and HAADF images of undoped $VO_2$ thin films, STEM-HAADF contrast analysis of the W clustering, HAADF images of 1 at.% W doped $VO_2$ thin film on sapphire and 2 at.% W doped $VO_2$ thin film on silicon substrate are available free of charge via the Internet at http://pubs.acs.org.

## AUTHOR INFORMATION

### Corresponding Author
* E-mail: alexandre.gloter@u-psud.fr

### Notes
The authors declare no competing financial interests.

## ACKNOWLEDGMENT
This study was supported by Shanghai Key Basic Research Project (09DJ1400200), NanOxyDesign ANR-10-BLAN-0814 program and the European Union Seventh Framework Programme (No. FP7/2007-2013) under Grant Agreement No. n312483 (ESTEEM2). The authors acknowledge Katia March and Maya Marinova-Atanassova for their help in sample preparation and STEM measurements.

# Supporting Information

# Direct Evidence of Tungsten Clustering in $W_{0.02}V_{0.98}O_2$ Thin Films and its Effect on the Metal-to-Insulator Transition

Xiaoyan Li,[†‡¶] Alexandre Gloter,[‡]* Alberto Zobelli,[‡] Hui Gu,[†] Xun Cao,[†] Ping Jin,[†] Christian Colliex[‡]

[†]State Key Laboratory of High Performance Ceramics and Superfine Microstructures, Shanghai Institute of Ceramics, Chinese Academy of Sciences, Shanghai 200050, China

[‡]Laboratoire de Physique des Solides, CNRS UMR 8502, Université Paris Sud 11, Orsay 91405, France

[¶]University of Chinese Academy of Sciences, Beijing 100049, China

Corresponding Author
* E-mail: alexandre.gloter@u-psud.fr; Tel: +33 1 69 15 53 71

# 1. Deposition process of the thin film

Epitaxial 2 at.% W doped $VO_2$ thin film were deposited on sapphire (001) surface by reactive magnetron sputtering. A vanadium metal disk containing 2% tungsten was used as the target. Ultra-pure argon (99.999%) and reactive oxygen gas were introduced through separated mass flow controllers to give an Ar flow rate of 40 sccm (standard cubic centimeter) and required oxygen flow ratios ($O_2$ flow rate/total flow rate). Plates (12 x8 x0.3 mm) of mirror-polishing sapphire (001) single crystal was applied in the epitaxial growth. The W content of the thin film is confirm by EDS (X-Max, Oxford. co) in a SEM (Supra55, Zeiss. co) to be around 2-2.5 at.% (see Figure.S1).

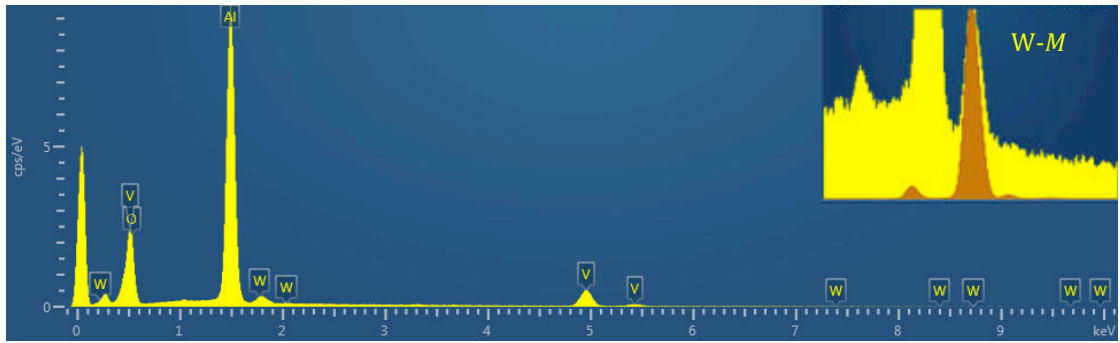

**Figure S1.** EDS spectrum of the $W_{0.02}V_{0.98}O_2$ thin film on (001) sapphire substrate.

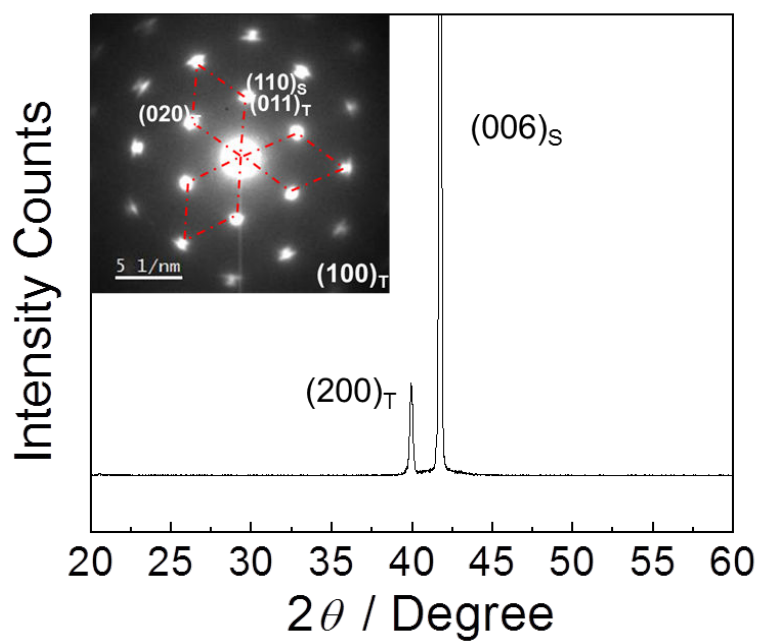

**Figure S2.** XRD $\theta$-$2\theta$ scan pattern of the $W_{0.02}V_{0.98}O_2$ thin film on (001) sapphire substrate taken at room temperature. Inset is its selected area electron diffraction (SAED) pattern taken from a TEM planar sample.

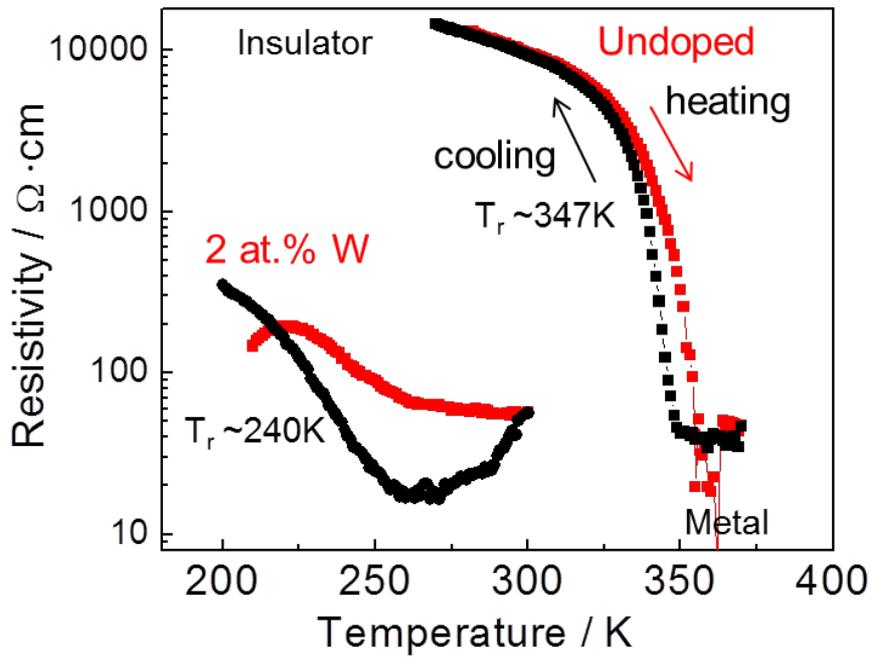

**Figure S3**. Resistivity curves as a function of the temperature for the 2 at.% W doped $VO_2$ thin film in comparison with an undoped $VO_2$ thin film which is deposited under the same condition. During the measurement, Ag was used for electrodes, which were deposited on the surface of the thin film. The contact was ohmic contact.

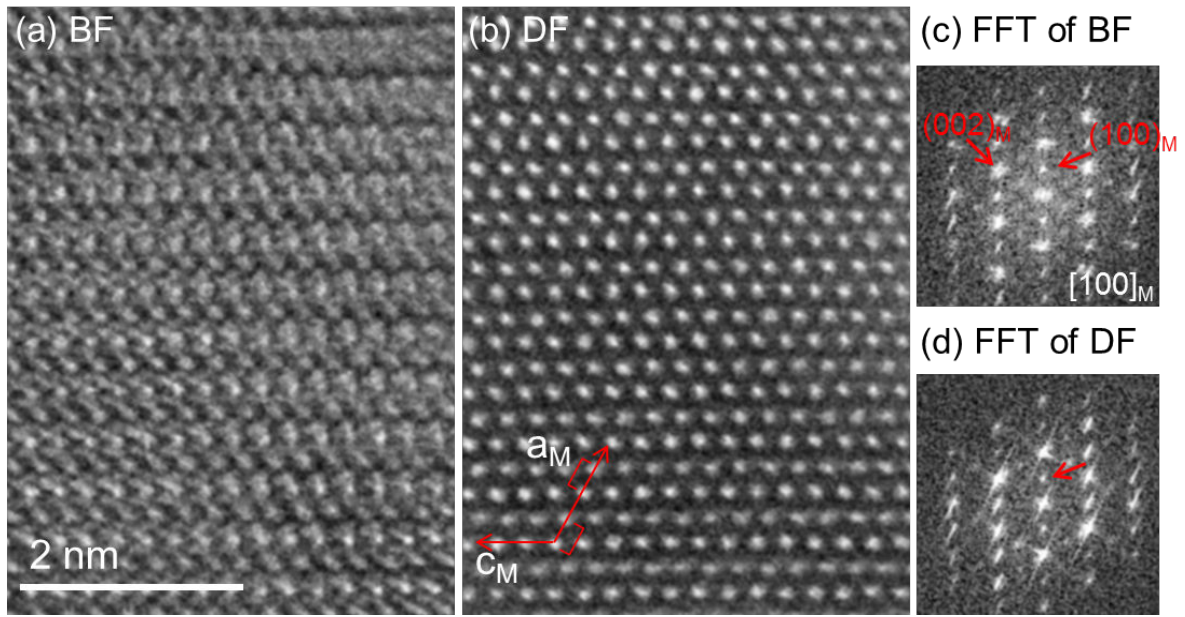

**Figure S4. (a)** BF and **(b)** HAADF images of the planar undoped $VO_2$ thin film TEM sample. **(c and d)** the FFTs for the BF and HAADF images respectively, within the $[010]_{M1}$ zone axis. When the image is taken in STEM in the room temperature, pure $VO_2$ is in the M1 phase, where the V-V pairing structures give a faint contrast modulation along the $[100]_{M1}$ direction in the HAADF image corresponding to the diffraction spot $(100)_{M1}$ in its FFT.

## 2. STEM-HAADF contrast analysis of W clustering

The Cs-corrected STEM-HAADF imaging techniques based on large convergence angles gives a reduced depth of field which has already been used to localize the dopant position in analysing the clustering of heavy elements in lighter matrix.[1-3] In the case of W doped $VO_2$, a similar trend is expected and observed in the Figure S5 where the relative HAADF intensities of W-containing columns are shown as a function of the W located depth. This HAADF simulation is realized in the QSTEM framework[4] with the parameters obtained from a typical experimental setup: convergence angle 30mrad, HAADF collection angles between 80-200 mrad, high voltage of 100 keV, residual Cs parameters of 0.001 mm, Scherzer focus of -2.4 nm. A $VO_2$ 44x3x5 supercell is used which is investigated along the $a_R$ axis and ca. 20 nm thick with several different W configurations: one single W, two-adjacent W and three-adjacent W located at six different depths.

The results confirm that at a given depth the intensity of the observed column is related to the number of W atoms. It also reveals that the HAADF intensity reaches an intense maximum when the W atoms are located at the defocus plane (-2.4 nm) with a typical depth range of around 3 nm which is in agreement with the geometrically estimated depth of field $\Delta_z = \lambda/\alpha^2 = 4$ nm. Thus, the neighboring atoms with similar enhanced intensity as observed in the STEM-HAADF images are at the similar depths suggesting short range atomic clustering. This is the case for the much brighter columns in Figure 3c, 4a and 5 that have been investigated for the W clustering preferences analysis, and some typical examples are encircled in these figures. The absolute quantification of the numbers of W atoms per column is not reported since it cannot be made unambiguous due to the overlap of the intensities for different W configurations located at different depths, notably for the less bright W-contained columns.

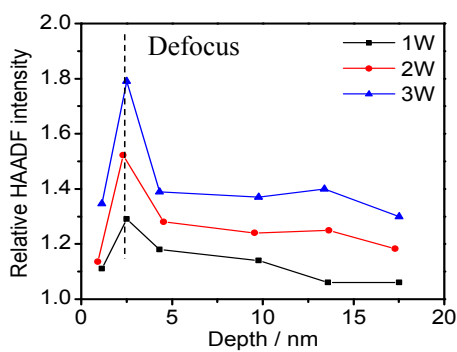

**Figure S5.** The relative HAADF intensity of the W-contained column as a function of the depth. The intensities have been normalized to the pure V columns.

### 3. Clustering of the W in the VO$_2$ at lower concentration

Figure S6a displays a STEM-HAADF image for the 1 at.% W doped VO$_2$ thin film on the (001) sapphire deposited in the same condition as for the 2 at.% W. In the large lateral box, a cluster can be found to start to form a local ordering along [010]$_R$. In the vertical box, a cluster appears with a modulation of intensity along the [001]$_R$ and the very bright columns indicating the presence of more than one W atoms. Others smaller W clusters can also be observed in the figure. In addition, the existence of some regions free of bright columns confirms that W atoms are heterogeneously distributed.

A schematic representation of one of the possible atomic model compatible with the contrast geometry observed in the lateral box is exhibited in Figure S6b. The structure is oriented the same as the grain in the HAADF image along [100]$_R$ zone axis. According to the clustering trend revealed by both experiments and *ab-initio* modeling in the main text, the distances between the W atoms are only along [010]$_R$, [011]$_R$ and [110]$_R$ directions corresponding to 4.55, 5.39 and 6.44 Å in the bulk R-VO$_2$, while the first neighbor positions along [001] or [½½½] are not present. In this model, seven W atoms are located in different (010)$_R$ planes at similar depth forming locally alternative ordering. Moreover, the three W atoms in the central part are located in the same (100)$_R$ plane forming a small cluster chain along [010]$_R$. The overall W distributions as seen along the [100]$_R$ direction is in accordance with the observed contrast geometry.

In Figure S6c, the same structural model has been rotated of ~90° around the *c* axis and is then now observed along the [010]$_R$ zone axis. This zone axis is crystallographically equivalent to the [100]$_R$ in the case of R-VO$_2$. It is interesting to note that when observed along this direction, the structural model is now more compatible with the as-observed vertical boxed cluster in the upper part of the Figure S6a. Indeed, the central column contains a very high W content (3-clustered W), which will result in a very bright column, surrounded by two lighter columns along the $c_R$ direction with less W content (1 W).

In these examples, we do not attempt to have quantitative model of every clusters but to demonstrate that the observed contrasts can be explained by the formation of small W clusters with the distances between the W atoms being mostly along [010]$_R$, [011]$_R$ and [110]$_R$.

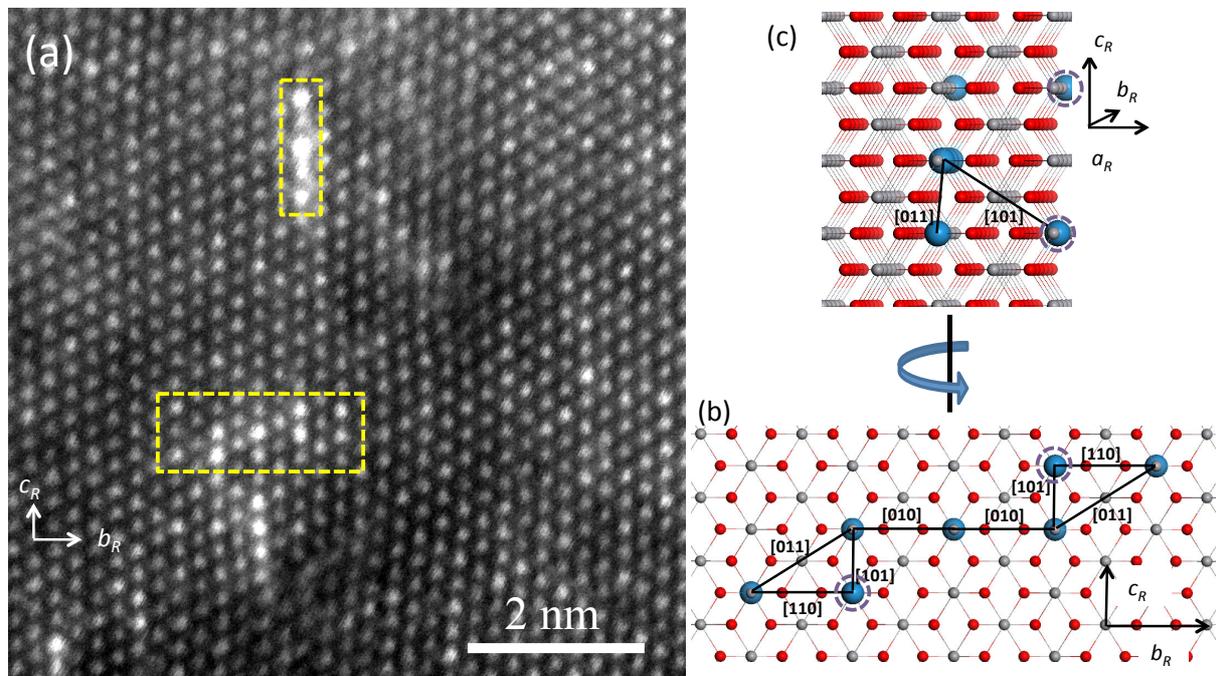

**Figure S6. (a)** STEM-HAADF image for the 1 at.% W doped VO$_2$ thin film on the (001) sapphire. **(b)** Schematic representation of atomic models compatible with the contrast geometry observed in the lateral box. The two circled W atoms in (b) are on the upper (100)$_R$ plane which located aside in (c), while the others are all on the lower (100)$_R$ plane. **(c)** The same structural models rotated ~90° around the $c_R$ axis. The two cluster along [011] and [101] noted in (c) correspond to the one at the left-down side of (b). The V and O atoms are grey and red respectively, while the large blue atoms are W and can then be seen whatever depth they are in the columns.

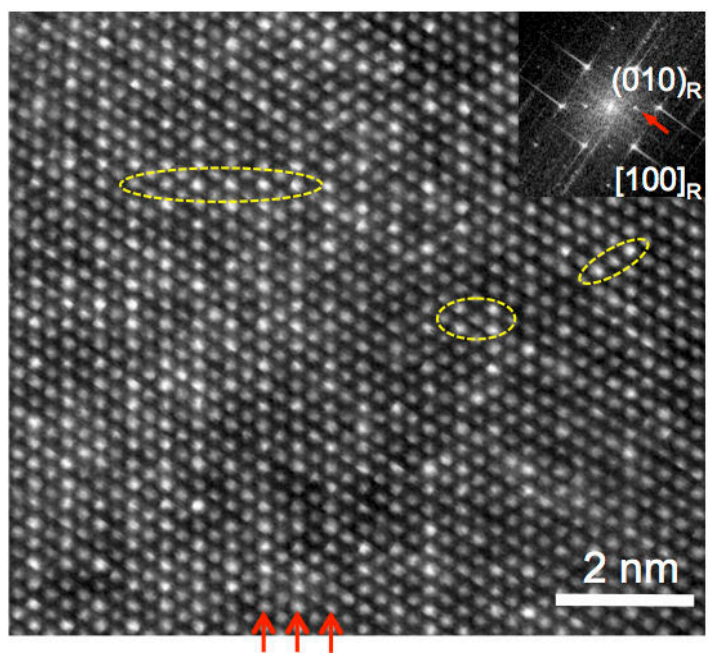

**Figure S7.** HAADF image of the 2 at.% W doped $VO_2$ thin film deposited on silicon substrate, the same W clustering features can be observed even in this non-epitaxial orientated thin film. Inset is the FFT of the HAADF image where the forbidden diffraction spot $(010)_R$ can be seen.